\documentclass[preprint,pre,showpacs,superscriptaddress]{revtex4}
\usepackage{graphicx}% Include figure files
\usepackage{dcolumn}% Align table columns on decimal point
\usepackage{bm}% bold math

\begin{document}

\title{Ground-state properties of the spin-1/2 Heisenberg-Ising bond alternating chain
       with Dzyaloshinskii-Moriya interaction}
\author{J. STRE\v{C}KA}
\affiliation{Department of Theoretical Physics and Astrophysics, Faculty of Science, P. J. \v{S}af\'{a}rik University,\\
Park Angelinum 9, 040 01 Ko\v{s}ice, Slovak Republic}
\author{L. G\'ALISOV\'A}
\affiliation{Department of Applied Mathematics, Faculty of Mechanical Engineering, Technical University,\\
Letn\'a 9, 042 00 Ko\v{s}ice, Slovak Republic}
\author{O. DERZHKO}
\affiliation{Institute for Condensed Matter Physics, National Academy of Sciences of Ukraine,\\
1 Svientsitskii Street, L'viv-11, 79011, Ukraine}

\begin{abstract}
Ground-state energy is exactly calculated for the spin-1/2 Heisenberg-Ising bond alternating chain
with the Dzyaloshinskii-Moriya interaction.
Under certain condition, which relates a strength of the Ising, Heisenberg and Dzyaloshinskii-Moriya interactions,
the ground-state energy exhibits an interesting nonanalytic behavior accompanied with a gapless excitation spectrum.
\end{abstract}
\pacs{05.50.+q; 05.70.Jk; 64.60.De; 75.10.Pq}

\maketitle

%\section{Introduction}

Quantum spin chains provide an excellent playground for theoretical studies of collective quantum phenomena
as they may exhibit numerous exotic ground states and quantum critical points \cite{matt93}.
The spin-1/2 Heisenberg-Ising bond alternating chain,
which has been originally invented by Lieb \textit{et al.} \cite{lieb61} and recently re-examined by Yao \textit{et al.} \cite{hyao02},
represents a valuable example of rigorously solved quantum spin chain.
The present work aims to provide a generalization of this simple but nontrivial quantum spin model
by taking into account the antisymmetric Dzyaloshinskii-Moriya interaction.

%\section{Heisenberg-Ising bond alternating chain}

Let us consider a bond alternating chain of $2N$ spins 1/2 with nearest-neighbor antiferromagnetic interactions,
which are alternatively of the Heisenberg and Ising type, respectively.
The total Hamiltonian of the model under consideration is given by
\begin{eqnarray}
H \!\! &=& \!\! \sum_{n=1}^{N} \Bigl[J_{\rm H} (s^x_{2n-1} s^x_{2n} +  s^y_{2n-1} s^y_{2n} +
\Delta s^z_{2n-1} s^z_{2n}) \nonumber \\
\!\! &+& \!\! D \left(s^x_{2n-1} s^y_{2n} - s^y_{2n-1} s^x_{2n} \right)
        + 2 J_{\rm I} s_{2n}^z s_{2n+1}^z \Bigr]\!,
\label{2.01}
\end{eqnarray}
where the parameter $J_{\rm H} (\Delta)$ denotes the XXZ Heisenberg interaction between $2n-1$ and $2n$ spins,
$\Delta$ is an anisotropy in this interaction,
and the parameter $D$ stands for the $z$ component of the antisymmetric Dzyaloshinskii-Moriya interaction present along the Heisenberg bonds.
Furthermore,
the term $2 J_{\rm I}$ denotes the Ising interaction between $2n$ and $2n+1$ spins
and the periodic boundary condition $s^{\alpha}_{2N+1} \equiv s^{\alpha}_1$ ($\alpha=x,y,z$) is imposed for convenience.

First,
let us eliminate from the Hamiltonian (\ref{2.01}) the Dzyaloshinskii-Moriya term after performing a spin coordinate transformation.
The spin rotation about the $z$-axis by the specific angle $\tan \varphi = D/J_{\rm H}$,
which is performed at all even sites $2n$ ($n=1,\ldots,N$),
\begin{eqnarray}
s^x_{2n}\to s^x_{2n}\cos\varphi+s^y_{2n}\sin\varphi,
\;
s^y_{2n}\to -s^x_{2n}\sin\varphi+s^y_{2n}\cos\varphi,
\nonumber
\end{eqnarray}
ensures a precise mapping equivalence between the Hamiltonian (\ref{2.01}) and the Hamiltonian
\begin{eqnarray}
H = \sum_{n=1}^{N}
\Bigl[\!\!\!\!\! && \!\!\!\!\!
\sqrt{J_{\rm H}^2 + D^2} \left(s^x_{2n-1} s^x_{2n} + s^y_{2n-1} s^y_{2n} \right) \nonumber \\
\!\!\! &+& \!\!\! J_{\rm H} \Delta s^z_{2n-1} s^z_{2n} + 2 J_{\rm I} s_{2n}^z s_{2n+1}^z \Bigr]\!. \label{2.03}
\end{eqnarray}
From here onward, one may closely follow the rigorous procedure developed in Refs. \cite{lieb61,hyao02}.
According to this,
the Hamiltonian (\ref{2.03}) is rewritten in terms of raising and lowering operators in the subspace where the ground state is,
and subsequently, the Jordan-Wigner transformation is applied
to express the relevant spin Hamiltonian as a bilinear form of Fermi operators.
The Fourier and Bogolyubov transformations are finally employed
to bring the Hamiltonian relevant for the ground-state properties into the diagonal form
\begin{eqnarray}
H = -\frac{N}{4} J_{\rm H} \Delta
+ \sum_{k} \Lambda_{k} \left(\beta^{\dagger}_{k} \beta_{k} - \frac{1}{2} \right),
\label{2.10}
\end{eqnarray}
where
\begin{eqnarray}
\Lambda_{k} = \sqrt{\left(\sqrt{J_{\rm H}^2 + D^2} + J_{\rm I} \right)^2
                  - 4 \sqrt{J_{\rm H}^2 + D^2} J_{\rm I} \cos^2 \frac{k}{2}}.
\label{2.11}
\end{eqnarray}
From Eqs. (\ref{2.10}) and (\ref{2.11}) one easily finds the exact result
for the ground-state energy of the antiferromagnetic spin-1/2 Heisenberg-Ising bond alternating chain (\ref{2.01})
for $N\to\infty$
\begin{eqnarray}
\frac{E_0}{N} = -\frac{1}{4} J_{\rm H} \Delta
                      - \frac{\sqrt{J_{\rm H}^2 + D^2} + J_{\rm I}}{\pi} {\rm{E}}(a),
\label{2.12}
\end{eqnarray}
where
${\rm{E}}(a)=\int_{0}^{\frac{\pi}{2}}{\rm{d}} \theta \sqrt{1-a^2 \sin^2 \theta}$
is the complete elliptic integral of the second kind with the modulus $a$,
\begin{eqnarray}
a^2 = \frac{4 \sqrt{J_{\rm H}^2 + D^2} J_{\rm I}}{\left(\sqrt{J_{\rm H}^2 + D^2}
                                                        + J_{\rm I} \right)^2} \geq 0.
\nonumber
\end{eqnarray}
Recall that the complete elliptic integral of the second kind
is a nonanalytic function of its modulus for $a^2 = 1-(a^{\prime})^2 \approx 1$,
i.e., ${\rm{E}}(a)-1\propto \ln a^{\prime} (a^{\prime})^2$.
The condition $a^2 = 1$ holds just if $J_{\rm I} = \sqrt{J_{\rm H}^2 + D^2}$
and hence, one may expect nonanalytic behavior of the ground-state energy (\ref{2.12}) under this special constraint,
which relates a strength of the Ising, Heisenberg and Dzyaloshinskii-Moriya interactions.

%\section{Results and Discussion}

Before proceeding to a more detailed discussion of the most interesting results,
it is worthy to mention that our exact results correctly reproduce
(in an absence of the Dzyaloshinskii-Moriya term)
the results previously reported by Lieb \textit{et al.} \cite{lieb61} for the isotropic version
and by Yao \textit{et al.} \cite{hyao02} for the anisotropic version
of the antiferromagnetic spin-1/2 Heisenberg-Ising bond alternating chain.
For simplicity,
our subsequent analysis will be restricted just to a particular case of the model with the isotropic Heisenberg interaction ($\Delta = 1$),
which exhibits all general features notwithstanding this limitation.

\begin{figure}[htb]
\includegraphics[width=0.52\linewidth]{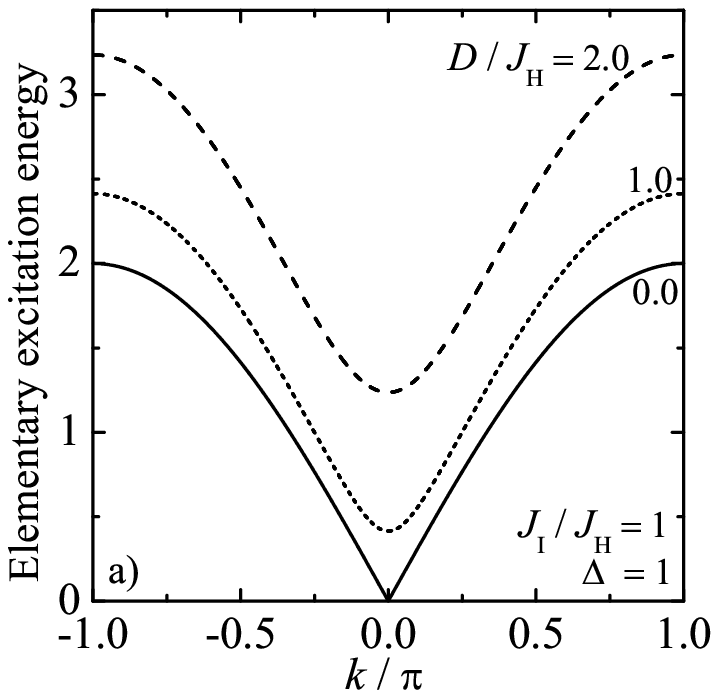}
\hspace{-0.6cm}
\includegraphics[width=0.52\linewidth]{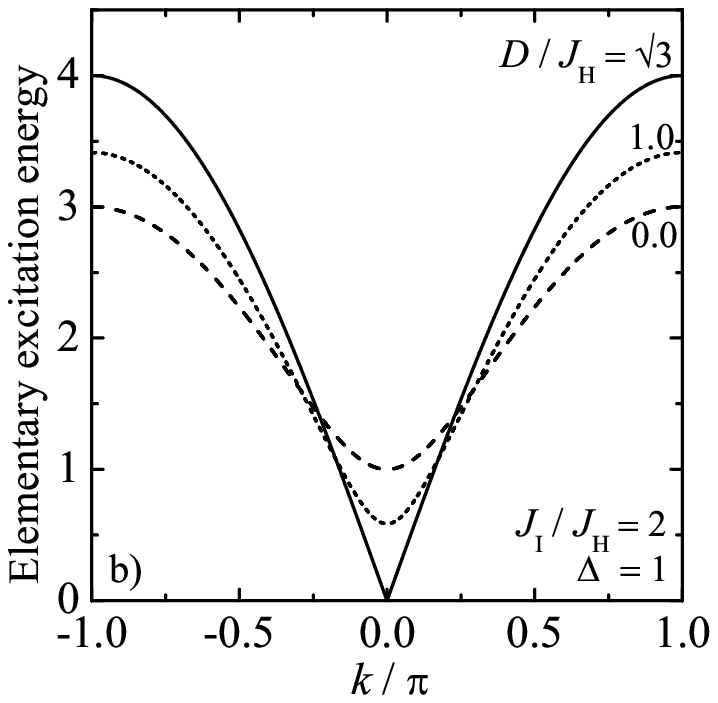}
\vspace{-1.1cm}
\caption{Elementary excitation spectrum for several values of the Dzyaloshinskii-Moriya term $D/J_{\rm H}$, $\Delta = 1$
and two different values of the ratio:
a) $J_{\rm I}/J_{\rm H} = 1$;
b) $J_{\rm I}/J_{\rm H} = 2$.}
\label{fig2}
\end{figure}

In Fig.~\ref{fig2}
we depict the elementary excitation energy spectrum $\Lambda_{k}$ calculated from Eq.~(\ref{2.11})
for two different values of the ratio $J_{\rm I}/J_{\rm H}$
and several values of the Dzyaloshinskii-Moriya anisotropy $D/J_{\rm H}$.
Generally, the excitations are gapped with exception of the particular cases
that satisfy the condition $J_{\rm I} = \sqrt{J_{\rm H}^2 + D^2}$.
The gapless excitation spectrum might be consequently found just
if $J_{\rm I}/J_{\rm H} \geq 1$,
which means that the Ising interaction must be at least twice
as large as the Heisenberg one. If $D/J_{\rm H} = 0$ is assumed, the system has gapless excitation spectrum for $J_{\rm I}/J_{\rm H} = 1$ in accordance with the previously published results  \cite{lieb61,hyao02}. Interestingly, the gapless excitation spectrum emerges at higher values
of the ratio $J_{\rm I}/J_{\rm H}$ regardless of the exchange anisotropy $\Delta$ whenever
the Dzyaloshinskii-Moriya anisotropy is raised from zero.

\begin{figure}[htb]
\vspace{2.2cm}
\begin{center}
\includegraphics[width=7.5cm]{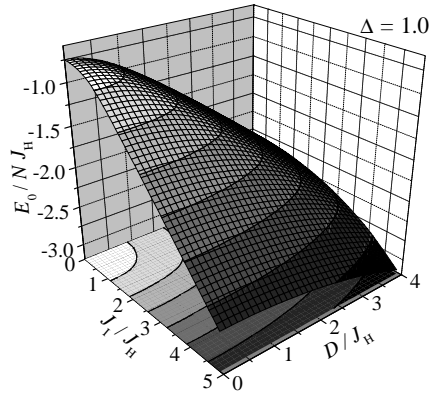}
\end{center}
\vspace{-2.8cm}
\caption{Ground-state energy as a function of the Dzyaloshinskii-Moriya anisotropy $D/J_{\rm H}$
and the interaction ratio $J_{\rm I}/J_{\rm H}$
for the anisotropy parameter $\Delta = 1$.}
\label{fig1}
\end{figure}

The three-dimensional plot of the ground-state energy (\ref{2.12}) is depicted in Fig.~\ref{fig1}
as a function of the ratio $J_{\rm I}/J_{\rm H}$ between the Ising and Heisenberg interaction,
as well as, a relative strength of the Dzyaloshinskii-Moriya anisotropy $D/J_{\rm H}$.
Referring to this plot,
the ground-state energy monotonically decreases upon strengthening the ratio $J_{\rm I}/J_{\rm H}$
and/or the Dzyaloshinskii-Moriya term $D/J_{\rm H}$.
In accordance with this statement,
the ground-state energy $E_0/N J_{\rm H} = -3/4$ of a system of the isolated Heisenberg dimers,
which is achieved in the limit $J_{\rm I}/J_{\rm H} \to 0$ and $D/J_{\rm H} \to 0$,
represents an upper bound for the ground-state energy.
Within the manifold $J_{\rm I} = \sqrt{J_{\rm H}^2 + D^2}$,
the ground-state energy exhibits a rather striking nonanalytic behavior.
Although this weak nonanalytic behavior cannot be seen from Fig.~\ref{fig1},
it should manifest itself in higher derivatives of the ground-state energy.

In the present work,
the ground-state properties of the spin-1/2 Heisenberg-Ising bond alternating chain with the Dzyaloshinskii-Moriya interaction
have been investigated using a series of exact (rotation, Jordan-Wigner, Fourier, Bogolyubov) transformations.
Exact results for the ground-state energy and elementary excitation spectrum
have been examined in relation with a strength of the ratio between the Ising and Heisenberg interaction, as well as, the Dzyaloshinskii-Moriya term.
The most interesting finding to emerge from our study closely relates to a remarkable nonanalytic behavior of the ground-state energy,
which is accompanied with the gapless excitation spectrum whenever the condition $J_{\rm I} = \sqrt{J_{\rm H}^2 + D^2}$ is met.

{\bf Acknowledgments}:
J.S., L.\v{C}., and O.D. thank the Abdus Salam International Centre for Theoretical Physics (Trieste, Italy) for hospitality during the School and Workshop
on Highly Frustrated Magnets and Strongly Correlated Systems: From Non-Perturbative Approaches to Experiments (2007), where the present study was launched.
J.S. and L.\v{C}. acknowledge financial support provided
under the grants VEGA 1/0128/08 and VEGA 1/0431/10.

\end{document}